# Dynamic effect of electron-number parity in metal nanoparticles

K. Son[1,2], D. Park[3,4], C. Lee[5], A. Lascialfari[6], S. H. Yoon[7], K. –Y. Choi[8], A. Reyes[9], J. Oh[10], M. Kim[10], F. Borsa[6], G. Schütz[1], Y. -G. Yoon[5]*, Z. H. Jang[11]*

[1]*Department of Modern Magnetic Systems, Max-Planck-Institute for Intelligent Systems, Stuttgart, 70569, Germany.*

[2]*Departement of Physics Education, Kongju National University, Gongju, 32588, Korea.*

[3]*Radwaste Technology Department, Korea Radioactive Waste Agency, Gyeongju 38062, Korea*

[4]*Department of Chemistry, Kookmin University, Seoul 02707, Korea*

[5]*Department of Physics, Chung-Ang University, Seoul 06974, Korea*

[6]*Departimento di Fisica, Università di Pavia, Pavia, Italy*

[7]*Department of Chemistry, Chung-Ang University, Seoul 06974, Korea*

[8]*Department of Physics, Sungkyunkwan University, Suwon 16419, Korea*

[9]*National High magnetic Field Laboratory, Tallahassee, FL 32310, USA*

[10]*Department of Materials Science and Engineering, Seoul National University, Seoul 08826, Korea*

[11]*Department of Physics, Kookmin University, Seoul 02707, Korea*

* Corresponding authors. Email: yyoon@cau.ac.kr, zeehoonj@kookmin.ac.kr

**Parity is a ubiquitous notion in science and serves as a fundamental principle for describing a physical system. Nanometre-scale metal objects are predicted to show dramatic differences in physical properties depending on the electron-number parity. However, the identification of the electron-number parity effects in real metal nanoparticles has been elusive because of the variations in various features of nanoparticles (size, surface morphology, crystallinity, etc.). Here we report the nuclear magnetic resonance (NMR) detection of the dynamic effect of the electron-number parity in silver nanoparticles. Field and temperature dependences of magnetization are in agreement with the calculations based on the electron-number-parity-dependent model with equal level spacing. With theoretical modeling of the NMR relaxation in silver nanoparticles, the nuclear spin-lattice relaxation rate is found to be proportional to the electron-number-parity-dependent susceptibility and to the temperature. This observation demonstrates the electron-number-parity-governed spin dynamics in silver nanoparticles.**

# Introduction

Over the past several decades, tremendous scientific and technological progress has been made in the area of metal nanoparticles research.[1-7] Thanks to their small-size, large surface-to-volume ratio, and discrete energy levels, metal nanoparticles show bizarre physical and chemical properties and hold potential applications in a wide range of fields such as catalysis, magnetic sensing, optical detection, medical diagnosis, etc. Singularly, size, shape, and composition dependence of their properties offer key principles for tailoring and gauging their specific physical or chemical performances and technological functionalities.[2-6] Nevertheless, our knowledge about metal nanoparticles is still limited or incomplete. For example, from the fundamental aspect, our understanding of the physics of the electron-number parity is very limited, even though the electron-number parity is a fundamental attribute that determines the physical properties of metal nanoparticles. The research on this topic can deepen our understanding of the parity effect and open new fields of applications of metal nanoparticles.

Starting in the 1930s, there has been an enormous amount of theoretical studies on the nanometre scale metal structures with unique physical properties.[1,8] Thanks to such efforts, theoretical modeling of nanometre-scale metal nanoparticle as a confined "free" electron system has revealed that thermodynamic properties differ markedly between systems with even and odd number of electrons [9]: in the case of odd parity system, magnetic susceptibility exhibits a Curie-law-like behavior at low temperatures whereas magnetic susceptibility for even parity system vanishes as the temperature is lowered toward zero kelvin (see Fig. 1) Also, alternation of the electron-number parity as a function of field is expected.[9] Since then, several pioneering attempts have been made to detect such parity effects.[10-13] As can be surmised from the rarity of the experimental evidence of the electron-number parity effect, their experimental verification remains challenging. It is known that the large spin-orbit coupling, non-uniformities in size, non-uniform energy level spacing, etc. can make the electron-number parity effect indistinct.[14, 15] Moreover, a non-uniform distribution of energy level spacing renders the calculation of thermodynamic quantities impractical in many cases. It was not until the mid-1990s that the above-mentioned theoretical predictions were confirmed by measuring the static thermodynamic properties of Pd clusters with a very uniform structure - the number of atoms in the core being the same in all clusters.[16] However, in typical metal nanoparticles, this uniformity is not guaranteed and both types of parity can coexist. The inevitably occurring parity mixing in real nanoparticles disrupts the separation between even- and odd-parity in the static thermodynamic measurements.

Alternatively, from the complete spin pairing in even-parity confinements (all the energy levels are occupied by two electrons with opposite spins) and the existence of the unpaired spins in odd-parity confinements, we infer that the spatiotemporal structure of electron spin correlations can be markedly different between an even- and an odd-parity confinements. (From now on, we will refer to this as the *dynamic electron-number parity effect*.) Therefore, unlike the static thermodynamic quantities, there is a possibility that intermingled dynamic responses of even- and odd-parity confinements can be separated in the time or the frequency domain by utilizing a dynamic probe. The nuclear magnetic resonance (NMR) technique is an excellent experimental tool to probe the dynamical response.

Herein, we report the observation of the dynamic electron-number parity effect in silver nanoparticles (AgNP) synthesized with an archetypical wet chemistry procedure with palmitic acid as the capping agent. $^1$H NMR measurements exhibit an unusual double nuclear magnetization relaxation, hinting at two distinct spin correlations. The $^1$H NMR nuclear spin-lattice relaxation rates (NSLRs) and the magnetization data are comprehensively analyzed with electron-number-parity dependent models, giving clear evidence for the existence of both *static* and *dynamic* electron-number parity effect in AgNP. The deduced value of level spacing suggests that the size of the confinement responsible for the electron-number parity effect is smaller than the nominal size of nanoparticles.

## Results

**Static electron-number parity effect**

We first examine the magnetization to check the static electron-number parity effect in AgNP specimen. As sketched in Fig. 1b and 1d, the field- and the temperature-dependent magnetization behavior is distinctly different for odd- and even-parity

In spite of the apparent paramagnetic-like behavior of isothermal magnetization (Fig. 2a), simultaneous fitting of the four isothermal $M_{exp}(T, H)s$ with the Langevin function could not be done with a common effective magnetic moment, meaning that the magnetism of AgNP is not simple paramagnetism. Instead, we resort to the equal level spacing model by Denton et. al.[9], which enables the calculation of the magnetization $m_{odd}(\delta; T, H)$ [$m_{even}(\delta; T, H)$] of the confinement with an odd [even] electron-number parity. Here, $\delta$ is the average value of level spacing of the discrete energy level of AgNP and $\delta$ is fixed to the value (196.58 K) obtained from the analysis of the NSLR for a consistent analysis of the measured $M_{exp}(T, H)$ data and the NSLR data (see Dynamic electron-number parity effect section). Noteworthy is that the average lever spacing of $\delta = 196.58$ K is much larger than the electronic Zeeman energy at the applied magnetic field of $\mu_0 H_{max} = 7$ T for our magnetization measurements, thereby the even-parity contribution to $M_{exp}(T, H)$ is negligible.[9] Thus, the total magnetization can be approximated as,

$$M_{exp}(T, H) = N_{odd} m_{odd}(\delta; T, H) + \chi_{bg} H, \tag{1}$$

where $N_{odd}$ is the number of the odd parity confinements per gram of AgNP sample and the last term contains the background susceptibility $\chi_{bg}$ arising from the capping material, etc. All four isothermal magnetization data were fitted simultaneously and the resultant fitting curves are in good agreement with the experimental data as shown in Fig. 2a. It is deduced from the fitting that $N_{odd}$ is $7.33 \times 10^{18}$ per gram of AgNP and $\chi_{bg}$ is $-1.02 \times 10^{-7}$ emu g$^{-1}$ Oe$^{-1}$. The background susceptibility is consistent with that of palmitic acid determined from independent measurement (see Supplementary Information).

Shown in Fig. 2b is the temperature-dependent magnetization measured at $\mu_0 H_{max} = 6.55$ T. The temperature-dependent magnetization is compared with the theoretical curve calculated using Eq. (1). In doing that, we used the same parameters obtained from fitting the isothermal $M_{exp}(T, H)$. As such, the coincidence between the theoretical curve and the experimental data gives credence to the employed fitting procedures. The simultaneous description of the field- and temperature-dependent $M_{exp}(T, H)$ using the common parameters provides evidence for the existence of the static electron-number parity effect in the AgNP sample.

**Dynamic electron-number parity effect**

In Fig. 3, we present $^1$H solid-state NMR spectra measured for powder samples of AgNP and capping material (palmitic acid). Almost all the measured spectra exhibit a Gaussian line shape irrespective of temperature and field. Considering its rigid molecular structure and in powder form, the Gaussian spectrum of palmitic acid is reasonable; in fact, for rigid solid with an approximately isotropic angular distribution of inter-nuclear dipolar couplings between protons, the proton NMR line shape is often approximated by a Gaussian function.[17]

On the other hand, there is an additional stronger source of local field at the proton sites of AgNP: the dipolar field of the electron magnetic moments in the AgNP core. The Gaussian spectrum of AgNP implies that the overall static local field at the proton sites has a Gaussian distribution. This is reasonable considering the randomness of the relative orientation/distance between the protons in the capping layer and the electron magnetic moments in the AgNP core.

As can be seen in Fig. 3a and 3b, the centers of the palmitic acid spectra and AgNP spectra coincide with the Larmor frequency of $^1$H NMR without noticeable shift. We also observed that the shifts of the Gaussian lines of the AgNP relative to the spectral lines of the palmitic acid are negligibly small. Thus, we note that the average value of the distribution of the local magnetic field at the proton sites in AgNP due to the electron magnetic moments in the AgNP core is negligibly small.

From Fig. 3, one can also infer that the linewidth of the spectra of palmitic acid and AgNP is temperature and/or field dependent. Investigation on $^1$H solid-state NMR spectra linewidth of palmitic acid and AgNP revealed that the temperature dependence of the linewidth of AgNP is qualitatively explainable in terms of the temperature dependence of the magnetization of the system with odd electron-number parity. Detailed discussion of the temperature dependence and field dependence of the linewidth is given in Supplementary Information.

The nuclear spin-lattice relaxation (NSR) behavior in AgNP shows compelling evidence for the dynamic electron-number-parity effect. Figure 4a exhibits the $^1$H NMR nuclear magnetization recovery curves, $M(t)$, of the AgNP and the capping material measured at $\mu_0 H = 1.299$ T and $T = 30$ K. Several salient features are found.

First, in contrast to a single stretched-exponential (SE) relaxation behavior observed in the capping material (palmitic acid), we observed a double SE relaxation behavior for AgNP. Since the proton has a nuclear spin of 1/2, the double SE relaxation is unexpected. The single SE observed in the NSR of the "capping material only" sample indisputably demonstrates that the double SE relaxation of AgNP is definitely due to the influence of the AgNP cores. This observation gives an indication that there are various sources of local field fluctuations at the proton sites of AgNP, i.e., two different types of confinements in the AgNP cores (even- and odd-parity confinements) and capping material itself.

Second, the double SE relaxation gradually turns to a single SE relaxation as the temperature is increased toward $T$ = 100 K as shown in the plot of $M(t)$ vs. $\tau T$ in Fig. 4b. ($\tau$ is the delay between saturation pulse sequence and monitoring pulse sequence.) Figure 4b shows that the slow-relaxing component remains almost the same while the fast-relaxing component gets slower as the temperature is raised.

From the double SE relaxation data of AgNP, two NSLRs are extracted by fitting the $M(t)$ data with the equation,

$$M(t) = M_0 + M_S\left[1 - \exp(t/T_{1S}^*)^\beta\right] + M_F\left[1 - \exp(t/T_{1F}^*)^\beta\right], \tag{2}$$

where $M_0$ is the baseline correction, $M_S(M_F)$ is the amplitude of the slow(fast)-relaxing component, $T_{1S}^*(T_{1F}^*)$ is the characteristic relaxation time for the slow(fast)-relaxing component, and $\beta$ is the common stretching exponent which is fixed to 0.87505 in the fitting. In Fig. 4c and 4d, the resulting NSLRs, $1/T_{1S}^*$ and $1/T_{1F}^*$, are divided by temperature, and plotted as functions of temperature for two different fields. The fast component $(T_{1F}^* T)^{-1}$ shows a divergent behavior as the temperature is lowered. Above $T$ = 100 K, $(T_{1F}^* T)^{-1}$ becomes almost temperature-independent and tends to merge with the slow component $(T_{1S}^* T)^{-1}$. On the other hand, $(T_{1S}^* T)^{-1}$ is almost constant in the whole temperature range of measurement. The $(T_1^* T)^{-1}$ of the capping material measured at 1.299 T is almost independent of temperature and comparable to the slow component $(T_{1s}^*T)^{-1}$ (Fig. 4c). This resemblance suggests that the slow component of local field fluctuations mainly arises from the local field fluctuations inherent to the capping material. Thus the very different behavior of the fast component below $T = 100$ K can be ascribed to the electron-number-parity dependent spin fluctuations in the AgNP core.

Stretched-exponential relaxation occurs when there is a distribution of $T_1$ and, for a fixed stretching exponent, the value of $T_1^*$ of a stretched exponential function is usually similar to or scales with the most probable value of $T_1$ in the distribution.[18] Therefore, from now on, we will use $T_1$ instead of $T_1^*$ for simplicity.

The NSLR, $1/T_1$, can be expressed as a Fourier transform of the correlation function of the transverse local field fluctuations at the nuclear sites.[19, 20]

$$\frac{1}{T_1} = \frac{\gamma_n^2}{2} \int \langle h_+(t) h_-(0) \rangle e^{i\omega_n t} dt, \qquad (3)$$

where $\langle h_+(t) h_-(0) \rangle$ is the autocorrelation function of the local field fluctuations, $\omega_n$ is the nuclear Larmor frequency, and $\gamma_n$ is the nuclear gyromagnetic ratio. The fluctuation of the local field is directly related to the fluctuation of the magnetic moments to which the nuclei are coupled via various interactions. Among them, the dipolar interaction is the dominant one in the investigated sample. Thus, the autocorrelation function of the local field fluctuation is directly related to the correlation function of the magnetic moment fluctuations via dipolar interaction. Our picture of magnetic moments, probing nuclei and coupling between them is schematically sketched in Fig. 5.

The correlation function of the magnetic moment fluctuations can be expressed as a product of the time-independent part and the time-dependent part. The time-independent part can be approximated as a product of the temperature and the static susceptibility.[21, 22] The time-dependent part is approximated as follows: given that $^1$H NMR of AgNP shows a Gaussian spectrum, it is reasonable to assume that the time dependence of the correlation function is of Gaussian type.[23, 24, 25] As a result, it is deduced that the NSLR, $1/T_1$, is proportional to the product of the temperature $T$ and the magnetic susceptibility $\chi$ as (see Supplementary Information for detail),

$$1/T_1 \propto T\chi \text{ and } 1/T_1 = \alpha T\chi. \tag{4}$$

We also devised a second model with the concept of spin diffusion in the hydrodynamic limit. In this model, it is assumed that the spin diffusion is restricted inside the nanometre-sized confinement. This second model also revealed that NSLR $(1/T_1)$ is proportional to $T\chi$, the same as the first model (see Supplementary Information for detail).

The magnetic moments responsible for the local field fluctuations are the electron spin moment in the AgNP cores and the magnetic moments in the background (capping material, etc.). Thus, the autocorrelation function of the local field fluctuation can be decomposed into two parts - one from the AgNP cores and the other from the background. The final equation for fitting the NSLR data is,

$$\frac{1}{T_1 T} = \alpha_{np}\chi_{np} + \alpha_{bg}\chi_{bg} = \alpha_{np}\left(\chi_{np} + \frac{\alpha_{bg}}{\alpha_{np}}\chi_{bg}\right) \equiv \alpha(\chi_{np} + \eta), \tag{5}$$

where $T_1$ is the proton spin-lattice relaxation time, $\chi_{np}$ is the magnetic susceptibility $(dM/dH)$ of the AgNP core, and $\chi_{bg}$ represents the average magnetic susceptibility stemming from all other materials in the AgNP specimen. $\alpha_{np}$ and $\alpha_{bg}$ are the proportionality constants for the AgNP core and the background, respectively. The parameter $\eta$ is defined as $\eta = (\alpha_{bg}/\alpha_{np})\chi_{bg}$ which accounts for the background contribution (see Supplementary Information for detail). With the equal level spacing formalism,[9] we calculate the magnetic susceptibilities $\chi_{np} =$ $\chi_{np}(\delta; T, H)$ for the even- and odd-number parity confinements. The calculated susceptibilities are used in the fitting of the measured NSLR data. By using Eq. (5), we fitted the two sets of $(T_{1F}^*T)^{-1}$ data simultaneously with the three common fitting parameters ($\delta, \alpha, \eta$) (see Fig. 4c and 4d). The odd-parity susceptibility is used in the fitting since the divergent behavior of $(T_{1F}^*T)^{-1}$ is very similar to that of the odd-parity susceptibility. It is reasonable because, in the even-parity cases, spin pairing leads to the cancellation of the magnetic moments. On the other hand, in the odd-parity cases, there is one unpaired electron in each confinement and the fluctuations of the unpaired electron give rise to dominant effects on the nuclear spin-lattice relaxation.

As is evident in Fig. 4c and 4d, the theoretical curves reproduce the experimental data well. The fitting parameters obtained from the fitting are $\delta = 196.58$ K, $\alpha = 3.74 \times 10^{25}$ Oe/(s·K·emu), and $\eta = -6.23 \times 10^{-27}$ emu/Oe. In Fig. 4c and 4d, we plot $1/T_1T$ vs. $T$ calculated with the even-parity susceptibility using the same parameters extracted from the odd-parity fitting. The calculated even-parity $1/T_1T$ lies below the slow component $(T_{1s}^*T)^{-1}$, indicating that the even-parity $T_1$ is longer than the intrinsic $T_1$ of the background.

**Discussion**

Now we discuss the physical relevance and ramification of the extracted fitting parameters. First, we start with the average energy level spacing $\delta$ of the discrete energy levels of the AgNP core. In a free electron model, with the bulk electron density of silver, the confinement volume $V$ and $\delta$ are related as [1],

$$V \text{ [cm}^3\text{]} = 1.45 \times 10^{-18}/\delta \text{ [K]}. \tag{6}$$

Using Eq. (6), we find that the energy level spacing of $\delta = 196.58$ K corresponds to spherical confinement with a diameter of 2.4 nm. This value is several times smaller than the average particle size of 7.31 nm deduced from the SEM image analysis (see Supplementary Information). Rather, it is found to be close to an average domain size of 3.25 nm obtained from XRD analysis (see Supplementary Information). This finding suggests that the confinement of electrons responsible for the electron-number parity effect occurs at the domain level, not throughout the whole particle. Despite the poor crystallinity of the silver core, the shape irregularity, and the difference in the electron density between bulk and nanoparticles, the similarity of the values of the domain size extracted from NMR and XRD is quite remarkable.

We must note that Eq. (6) is based on a free electron model. In reality, the electrons in the AgNP core are not completely free. The very poor crystallinity of the AgNP core puts it in a category of an amorphous material and the electrons in AgNP are confined in the nanometre scale region. Therefore, the physical properties of electrons in the AgNP core will differ from those of bulk crystalline silver or free electrons, and thus, it is highly possible that its $V$-$\delta$ relationship deviates from Eq. (6).

The electronic energy level spacing in nanometre-scale confinement is affected by the variations in the physical shape of the confinement, surface structure, irregular crystallinity, etc. As a consequence, an energy level structure varies from confinement to confinement. Thus, in calculating physically observable quantities, the variation should be treated statistically.[1, 9, 26] We also would like to point out that, in our study, we could reproduce the experimental data without such statistical processing. When the temperature is low enough compared to the level spacing between the Fermi level and the first excited state level ($k_B T < \delta$), an equal level spacing scheme is quite a good approximation even without rigorous statistical processing since the populations of higher excited states are negligibly small and, thus, the structure of the higher excited states is not important. (We stress that the notable changes of the static and dynamical effects are observed at temperatures below 100 K, which lies well below $\delta = 196.58$ K.)

Now, we proceed with the physical interpretation of $\alpha$ and $\eta$. We could derive the expression of $\alpha$ by resorting to two different models (or approximations) of the correlation functions mentioned before. With the expression and from the fitted value of $\alpha$, the average distance between the protons and the electrons in AgNP is estimated to be ~ 2.5 nm (see Supplementary Information for detail). This value agrees with the morphology of the samples. We further recall that the local field at the $^1$H nuclei sites in AgNP is induced by the magnetic moments in the AgNP. Since the capping material is diamagnetic in nature, its net moment under an external magnetic field is anti-aligned to the electronic moments in the silver core. It results in a diminished effective magnetic field at the proton sites and thus, less fluctuations of the field. This gives a rationale for the negative value of $\eta$ in Eq. (5).

In summary, we have demonstrated electron-number-parity-dependent spin dynamics with $^1$H NMR spectroscopy in a silver nanoparticle system. We have shown that the NSLRs provide a powerful way to separate odd-parity physics from even-parity physics in nanometre-scale systems such that $1/T_1T$ is proportional to the parity-dependent susceptibility, allowing the detection of the dynamic electron-number parity effect. Our work opens new horizons for an in-depth understanding of electron-number parity effects in nanoparticles.

## Acknowledgments

## Author contributions

K. S. and G. S. performed SQUID measurement. K. S. and Z. H. J. carried out magnetization analysis. K. S. and G. S. performed XPS characterization. D. P. and S. H. Y. performed material synthesis and preliminary characterization (TEM, SEM, ICP-AES, TGA, XRD). J. O. and M. K. performed HRTEM measurement. C. L., K. Y. C., A. R., and Z. H. J. performed NMR experiment. A. L., F. B., Y.-G. Y. and Z. H. J. carried out analysis of the NMR data. A. L., F. B., Y.-G. Y., and Z. H. J. developed theoretical interpretation of the results. All authors discussed the results and participated in writing manuscript, and commented on the manuscript.

## Competing interests

Authors declare no competing interests.

## Additional information

**Supplementary Information** is available for this paper

**Correspondence and requests for materials** should be addressed to Z. H. J.

## Methods

### Synthesis

The Ag nanoparticles (AgNP) were synthesized by modifying the method reported in the literature.[27] $AgNO_3$ (purity > 99.5 %), toluene (purity > 99.5 %), trimethylamine (purity = 98 %), and palmitic acid were purchased from Sigma-Aldrich and used as received. Briefly, palmitic acid (5.6 g) was dissolved in triethylamine (40 mL) followed by the addition of $AgNO_3$ (3.6 g). After 10 min stirring, the solution changed to white slurry which was refluxed at 80 °C for 2 h. The synthesized Ag nanoparticles were purified three times by addition of acetone (35 mL) followed by centrifugation at 4000 rpm for 10 min. The as-synthesized Ag nanoparticles were dispersed in hexane (0.5 wt %, brown).

### Morphology and crystallinity characterization

**TEM/SEM** Samples for Transmission Electron Microscope (TEM) analysis were prepared by dropping Ag nanoparticles dispersed in hexane (0.5 wt %) on an amorphous carbon-coated copper grid. The images were obtained with JEOL JEM-2100F (200 keV). High-Resolution TEM observations were conducted using a Cs-corrected TEM operated at 300 kV (Themis Z S/TEM, Thermo Fisher Scientific). To take the Scanning Electron Microscope (SEM) images of the Ag nanoparticles, the particles were mounted on specimen stubs with double-sided adhesive carbon tape and coated with Au in a sputter coater for 2 min to avoid charging. SEM images were obtained with Hitachi FE-SEM S-4800.

**X-ray Diffraction** The X-ray diffraction pattern of the Ag nanoparticles was obtained with X-ray diffractometer (XRD; Bruker D8 Focus) using Cu Kα radiation.

**Impurity and Composition characterization**

**TGA** Thermal Gravimetric Analysis (TGA) was performed with TA instruments STA-1500 to quantitatively determine the amounts of the capping material. The heating rate was 10 ˚C/min and the analysis was performed under $N_2$ atmosphere.

**ICP-AES** The Inductively Coupled Plasma (ICP)-Atomic Emission Spectrometry (AES) measurements were performed with thermo elemental iris intrepid ICP-AES spectrometer system. Samples for ICP-AES analysis were prepared by dissolving Ag nanoparticles (0.200 g) in nitric acid (10 mL).

**XPS** The X-ray Photoemission Spectroscopy (XPS) was performed to analyse elemental composition and to check impurity with Thermo VG Theta Probe system (Thermo Fisher Scientific, USA.) The system employs monochromatic Al Kα radiation ($hv$ = 1486.68 eV) produced with an electrical power of 100 W. The base pressure was $8 \times 10^{-8}$ mbar. The X-ray spot size was about 400 μm in diameter.

**Magnetic property characterization**

The magnetic properties of the silver nanoparticles (AgNP) were investigated using a commercial superconducting quantum interference magnetometer (SQUID, Quantum Design MPMS-XL, USA). Field-dependent magnetization was measured at various temperatures by varying the field in the range $0 \leq T$  $\mu_0 H \leq 7$ T. Temperature-dependent magnetization was investigated at 6.55 T. Since overall signal of the sample is not significantly large, extra care has been exercised during the preparation and the measurement process to avoid sample contamination, e. g. all the metallic tools were excluded in the sample preparation procedures not to contaminate sample with the magnetic impurities.

**Solid-state nuclear magnetic resonance (NMR) experiment**

$^1$H NMR was performed at NHMFL at Tallahassee, FL. U.S.A. utilizing user facility (wide band solid-state NMR - 12 T sweepable superconducting magnet, home-made VTI (Variable Temperature Insert) and in-house developed pulsed RF FFT NMR spectrometer).

**$^1$H NMR spectrum** Spectra of AgNP and capping material were obtained by performing FFT (Fast Fourier Transform) on half of the Hahn echo. The length of $\pi/2$ pulse was in the range from 1.15 to 7 microseconds. The spectral width of RF pulses is confirmed to be always larger than the linewidth of the AgNP spectra.

**$^1$H NMR spin-spin relaxation ($T_2$)** Spin-spin relaxation time $T_2$'s of AgNP and capping material were measured as functions of temperature at 1.299 T (both AgNP and capping material) and 3.0 T (AgNP only) with the conventional Hahn echo sequence ($\pi/2 - \pi$). The relative phase between the first and the second pulses has been varied according to the 16 phase cycling scheme to reduce the distortion due to the ringing.[28] $\pi/2$ pulse length ranged from 1.15 to 7 microseconds and single exponential relaxation behavior was observed in all the measurements of the AgNP and the capping material (palmitic acid).

**$^1$H NMR nuclear spin-lattice relaxation ($T_1$)** $^1$H NMR nuclear spin-lattice relaxation ($T_1$) behavior was investigated at two fields (1.299 T and 3.0 T) as functions of temperature. The saturation recovery sequence was utilized in all the measurements with ($\pi/2$) pulse length

ranging from 1.15 to 7 microseconds depending on the tuning and matching conditions of the RF probe system. Nuclear magnetization recovery data were obtained by recording echo height as functions of delay between the saturation pulse sequence and the monitoring pulse sequence. Saturation pulse sequence is composed of one or two (π/2) pulses of which the separation is longer than $T_2$ and much shorter than $T_1$ measured at the same field and temperature conditions.

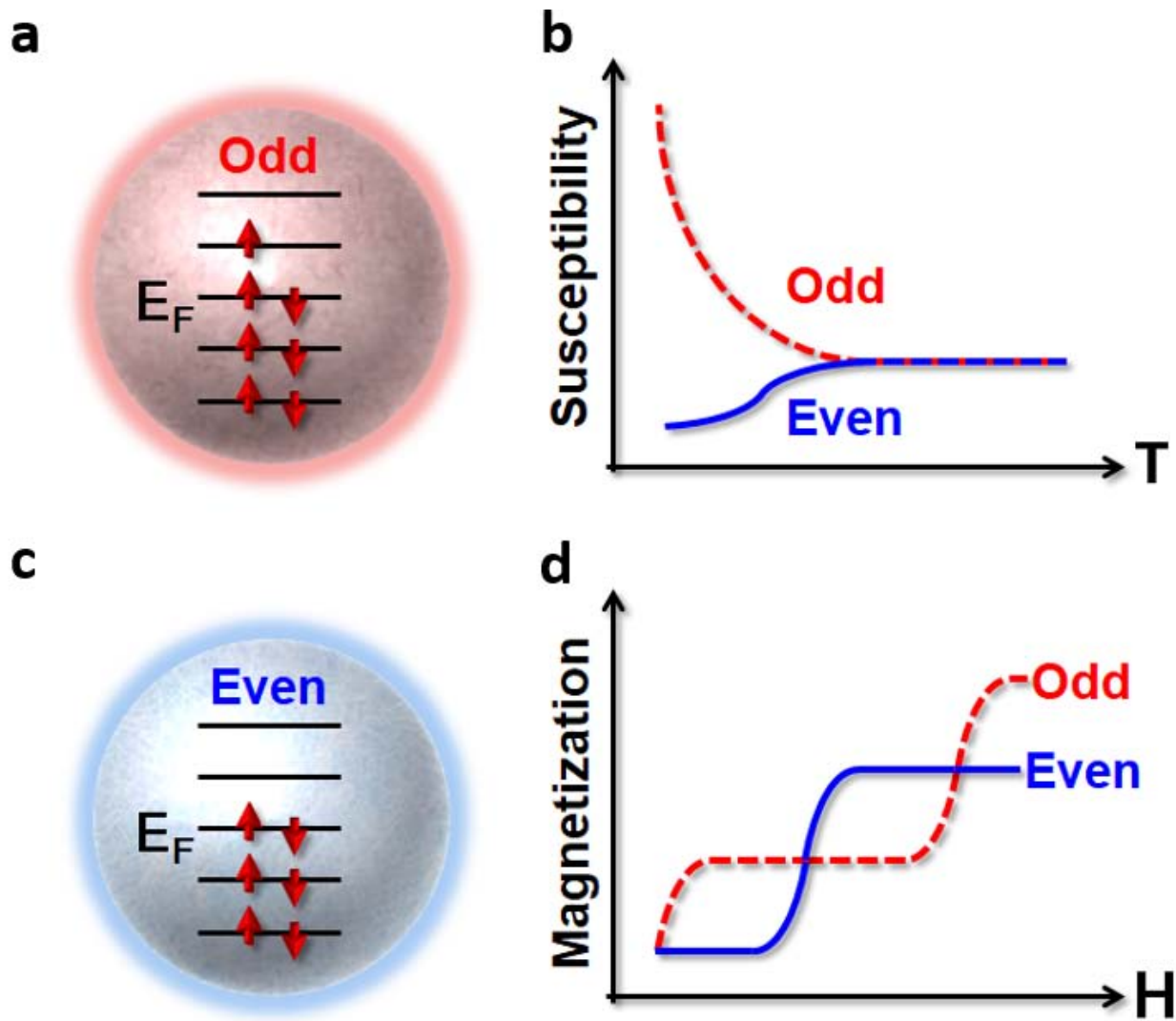

**Fig. 1. Schematic sketch of electron-number parity dependent magnetization and magnetic susceptibility. a**, **c**, Ground state electronic configurations. If an odd number of electrons are confined together (odd number parity, **a**), discrete energy levels up to the Fermi level are filled with two electrons (spin up and spin down pair) and the level just above the Fermi level is occupied by one unpaired electron. A divergent behavior of the temperature-dependent magnetic susceptibility (red dashed line in **b**) and a steep initial increase followed by a plateau in the field-dependent magnetization (red dashed line in **d**) originate from the unpaired electrons. If an even number of electrons are confined together (even number parity, **c**), all the energy levels up to the Fermi level are filled with two electrons and all the electrons are paired. The suppression of a temperature-dependent susceptibility (blue line in **b**) and an almost field-independent magnetization plateau at low field (blue line in **d**) result from the paired electron spins. It is worthwhile to note that the field-dependent magnetization alternates between odd- and even-parity behaviors at specific fields (spin-flip transition) whose field strength corresponds to half of the level spacing.[1,9] Similar effect occurs in the field-dependent magnetic susceptibility.

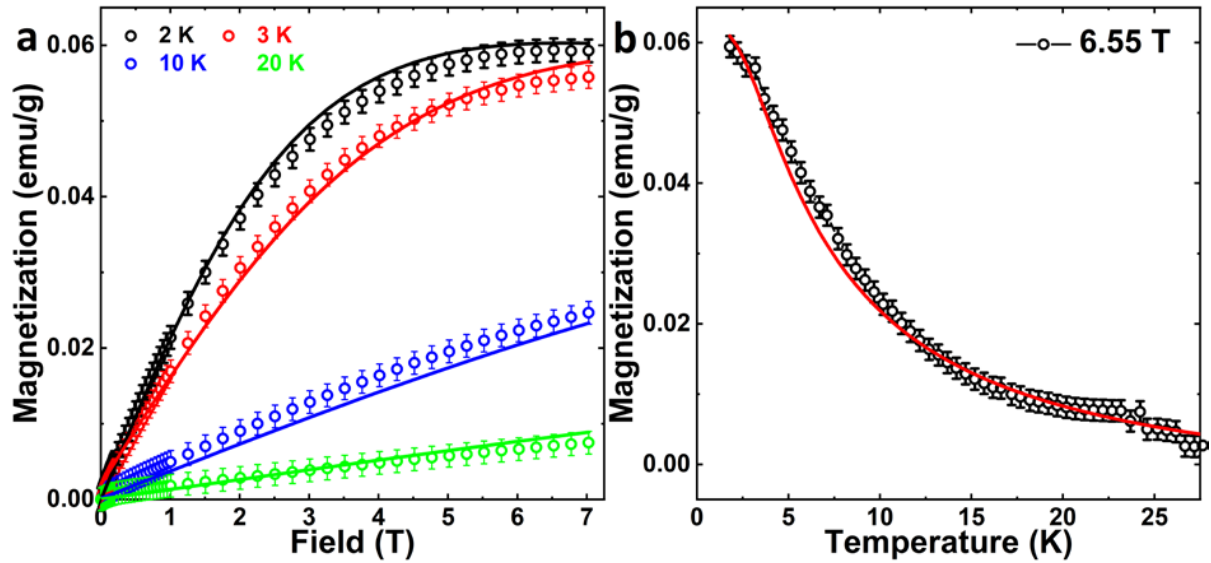

**Fig. 2. Electron-number parity effect in magnetization. a**, Isothermal field-dependent magnetization curves measured at four different temperatures (T = 2, 3, 10, 20 K). All the data are in agreement with the theoretical curves (solid lines) calculated with the model described in the main text. **b**, Temperature-dependent magnetization measured at 6.55 T. The red solid line represents the theoretical curve calculated using the same parameters deduced from the field-dependent isothermal magnetization fitting. The good agreement between the experimental and calculated magnetization provides independent confirmation of the model adopted in the analysis.

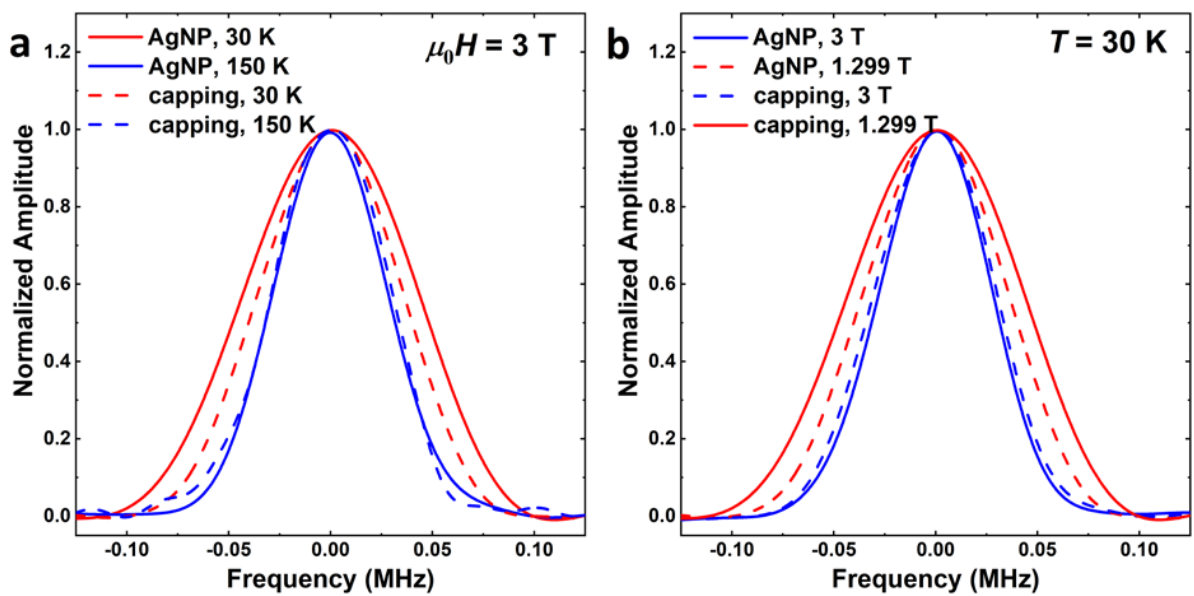

**Fig. 3. ¹H NMR spectra of silver nanoparticle. a**, Temperature dependent 1H NMR spectra of the AgNP and palmitic acid measured at 3.0 T, showing a Gaussian spectral shape and the increase of linewidth (FWHM) as the temperature is lowered. It is shown that the linewidth and shape of the spectra of the AgNP and palmitic acid are very similar to each other at high temperatures. **b**, Field dependent spectra of the AgNP and palmitic acid measured at 30 K. The spectral shape (Gaussian) is independent of the field, but the linewidth (FWHM) of the 3.0 T spectrum is larger than that of the 1.299 T spectrum, showing that the linewidth is field dependent. The spectra of palmitic acid from the separate measurements are also shown for comparison. It is noted that the scales of x-axis are expressed relative to the ¹H Larmor frequency in both graphs.

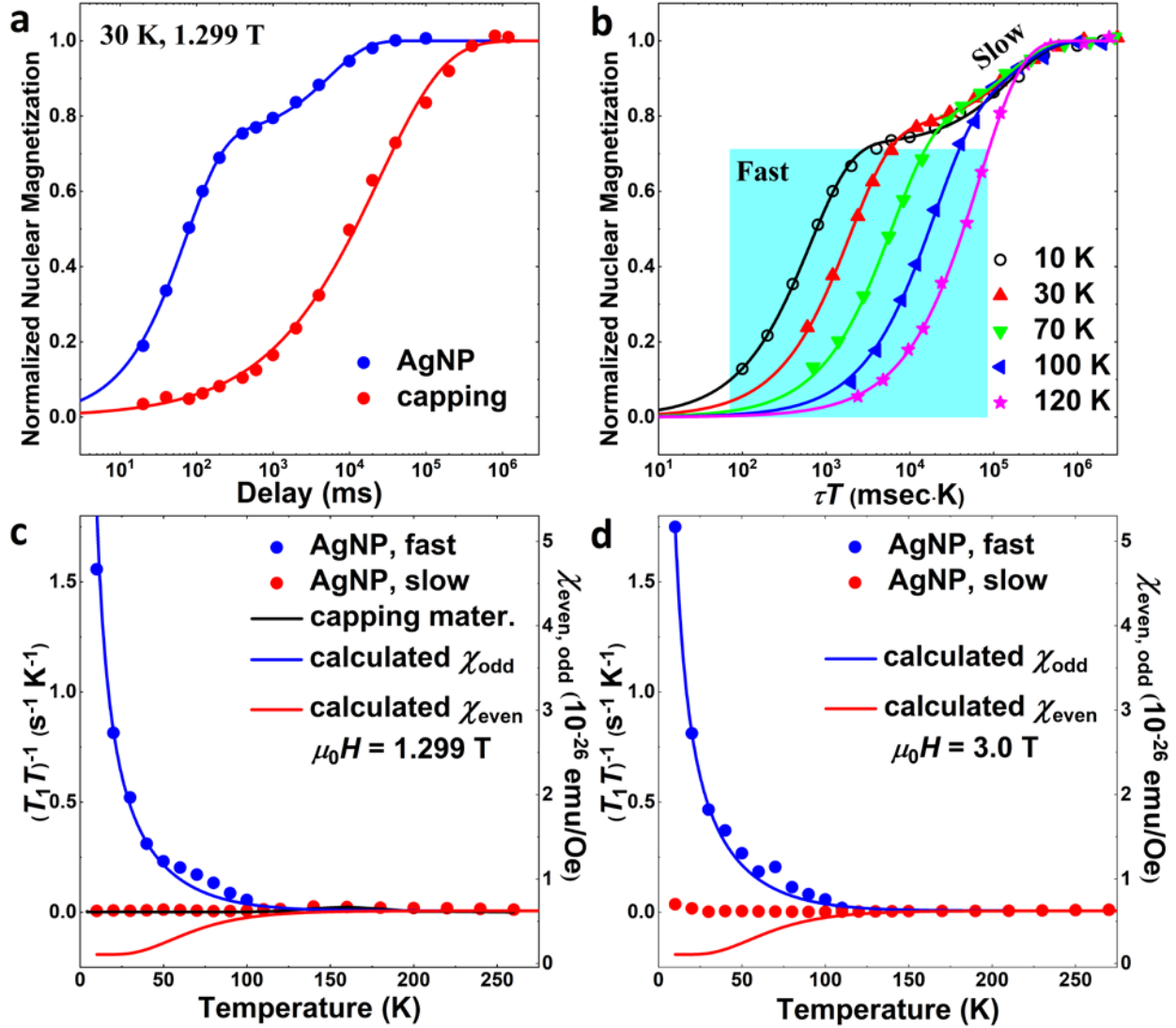

**Fig. 4. Electron-number parity dependent spin dynamics measured by nuclear spin-lattice relaxation of $^1$H NMR. a**, $^1$H NMR nuclear magnetization recovery curves of AgNP and capping material measured at $\mu_0 H = 1.299$ T and $T = 30$ K. Solid lines are fits to single stretched-exponential (capping material) and double stretched-exponential (AgNP). A double-step structure in the recovery of AgNP corroborates double stretched-exponential relaxation which is in contrast to the single stretched-exponential relaxation of the capping material. **b**, $^1$H nuclear magnetization versus $\tau T$ clearly shows that the fast component $1/T_1 T$ is strongly temperature-dependent. In contrast, the slow-component varies little with temperature. **c, d**, $1/T_1 T$ vs. $T$ of AgNP and capping material measured at 1.299 T and 3.0 T. For both fields, the fast component $1/T_1 T$ (blue circles) is divergent as the temperature is lowered below 100 K and almost constant above 100 K. The behavior is reminiscent of the odd parity susceptibility.[9] The slow component $1/T_1 T$ (red circles) and the $1/T_1 T$ of capping material (black solid line) are almost constant and coincide with each other over a whole temperature range (see **c**), suggesting that the slow component spin dynamics is dominated by the local field fluctuations inherent to capping material. The fast component $1/T_1 T$ data are successfully fitted with Eq. (5) using the electron-number-parity dependent susceptibility obtained from the equal level spacing model[9] (see blue solid lines).)

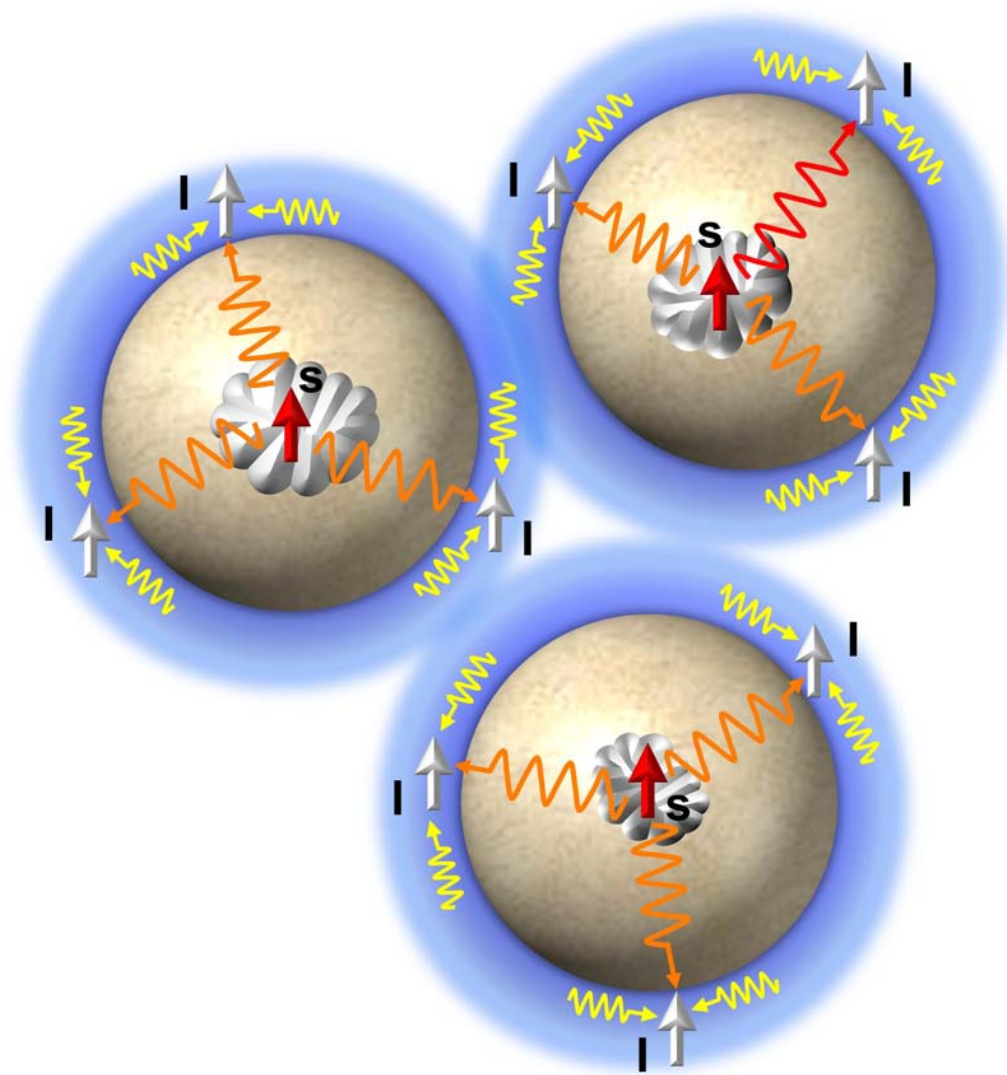

**Fig. 5. Schematic view of NMR detection of the spin dynamics in AgNP.** The unpaired electronic spins (**s**) in the silver cores are indicated by red arrows. Peach colored areas represent the silver core and shaded gray areas denote domains with odd electron-number parity. The blue circular ring regions are the capping material of AgNP. Light gray arrows are $^1$H nuclei (**I**) existing in the capping material. The local field at $^1$H nuclei (**I**) is composed of a dipolar field (orange wavy line) arising from the electron spins in the silver core and a dipolar field (yellow wavy line) arising from moments in the capping material.

# Supplementary Information

**Dynamic effect of electron-number parity in metal nanoparticles**

K. Son, D. Park, C. Lee, A. Lascialfari, S. H. Yoon, K. –Y. Choi, A. Reyes, J. Oh, M. Kim, F. Borsa, G. Schütz, Y. -G. Yoon*, Z. H. Jang*

*Correspondence to: yyoon@cau.ac.kr; zeehoonj@kookmin.ac.kr

# I. Morphology and crystallinity characterization

## I.1. Scanning Electron Microscope (SEM)

SEM image of the AgNP sample shows silver nanoparticles with a very regular spherical shape (Fig. S1). A particle size distribution (Fig. S1 inset) extracted from the SEM image is fitted with the Gaussian function. The fitting deduced that the average particle size and width of the size distribution are 7.31 nm and 2.31 nm, respectively. Interestingly, although the major portion of the nanoparticles is randomly packed, we can clearly identify a regular close-packed region with a narrower size distribution (standard deviation of the size distribution ~ 0.3 nm), seen in the upper left portion of the SEM image (Fig. S1), In the sample investigated for the magnetization and spin dynamics (NMR study), the random packing portion and close packing portion coexist.

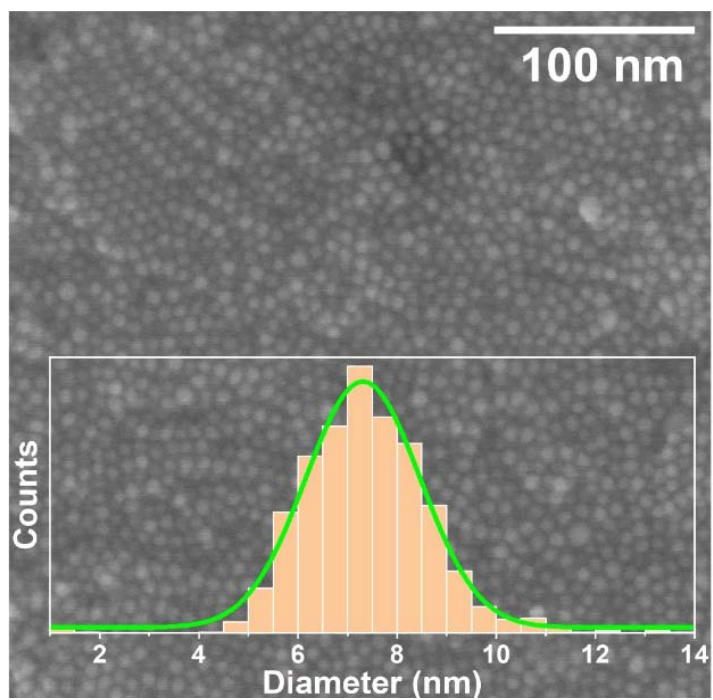

**Fig. S1: SEM characterization of AgNP.** SEM image of solidified AgNP specimen. The upper left portion shows a hexagonally close packed region where the particle size distribution is very narrow (standard deviation of the size distribution ~ 0.3 nm). AgNPs in regular spherical shape are observed. (inset; particle size distribution histogram obtained from SEM image of AgNP. The distribution is analyzed with Gaussian function and the average particle size and the width of the distribution are found to be 7.31 nm and 2.31 nm, respectively.)

## I.2. Transmission Electron Microscope (TEM)

Crystalline morphology of the AgNP was investigated with TEM (Fig. S2a and S2b). A whole particle single-crystallinity is absent but polycrystalline morphology is observed in all the particles, ascertaining a multi-domain nature. (Fig. S2a) Similarly, the High-Resolution TEM (HR-TEM) image of typical AgNP in Fig. S2b lacks a whole particle single-crystal morphology, consistent with the existence of multiple single-crystal domains.

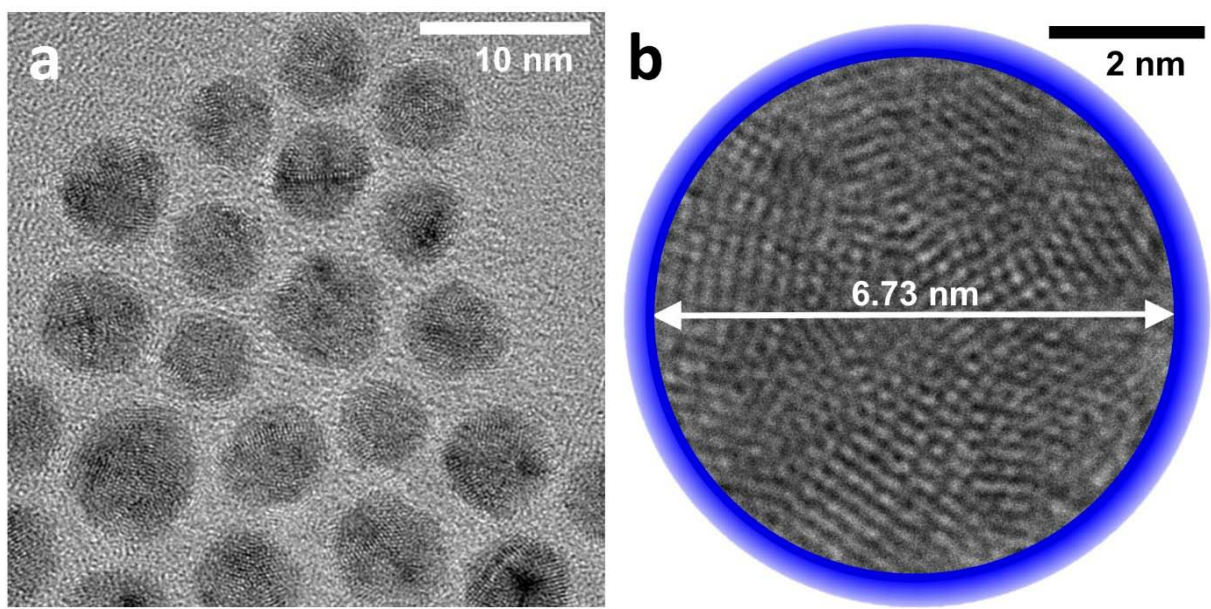

**Fig. S2: TEM characterization of AgNP. a**, TEM image of multiple particles showing polycrystalline morphology, confirming that all the Ag nanoparticles are composed of multiple crystalline domains. **b**, HR-TEM image of a typical AgNP. The outer blue circle represents the capping material surrounding the AgNP core. It is clear that whole particle crystallinity is lacking and a single particle is composed of multiple single-crystal domains.

## I.3. X-Ray Diffraction (XRD)

The existence of multiple single-crystal domains in a single particle is also confirmed by XRD results shown in Fig. S3. We could identify several characteristic peaks at specific positions of face-centered cubic (fcc) crystal structure of Ag(ICSD # 64706) but the peaks are broadened due to the finite size of the domains in particles. The size of the domains in Ag nanoparticles was calculated using FWHM (Full Width at Half Maximum) values of each diffraction peaks and the Scherrer formula,

$$D = \lambda/(\beta \cos\theta),$$

where $D$ is the average domain size, $\beta$ is FWHM in radian and $\theta$ is the Bragg angle. ($\lambda = 1.54$ Å for Cu K$\alpha$ radiation) Thus obtained domain size of the Ag nanoparticles is 3.25 nm, indicating that the Ag nanoparticles are composed of multiple domains. It should be noted that the estimated size from the broadened XRD peaks is not the nominal size of the nanoparticles but the average size of the single crystalline domains in the nanoparticles.

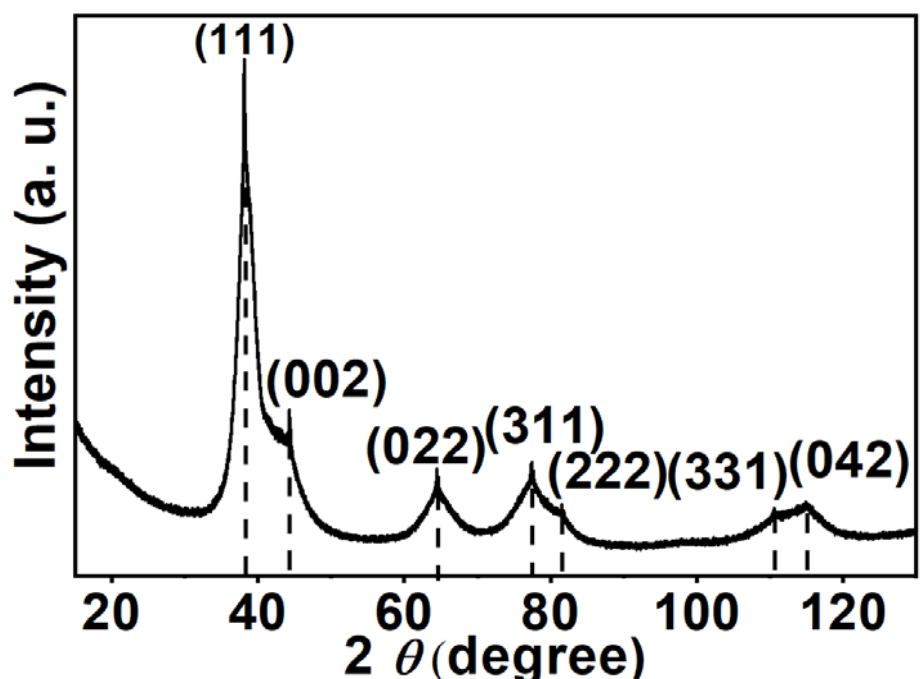

**Fig. S3: XRD characterization of AgNP.** X-ray Diffraction pattern of the AgNPs. The peak positions agree well with those of crystalline silver. The analysis of the broadened XRD peak revealed that the average size of a single crystal domain is 3.25 nm.

## II. Impurity and composition characterization

### II.1. Thermal Gravimetric Analysis (TGA)

TGA analysis revealed that the amount of the capping material (palmitic acid) is 18 wt %.

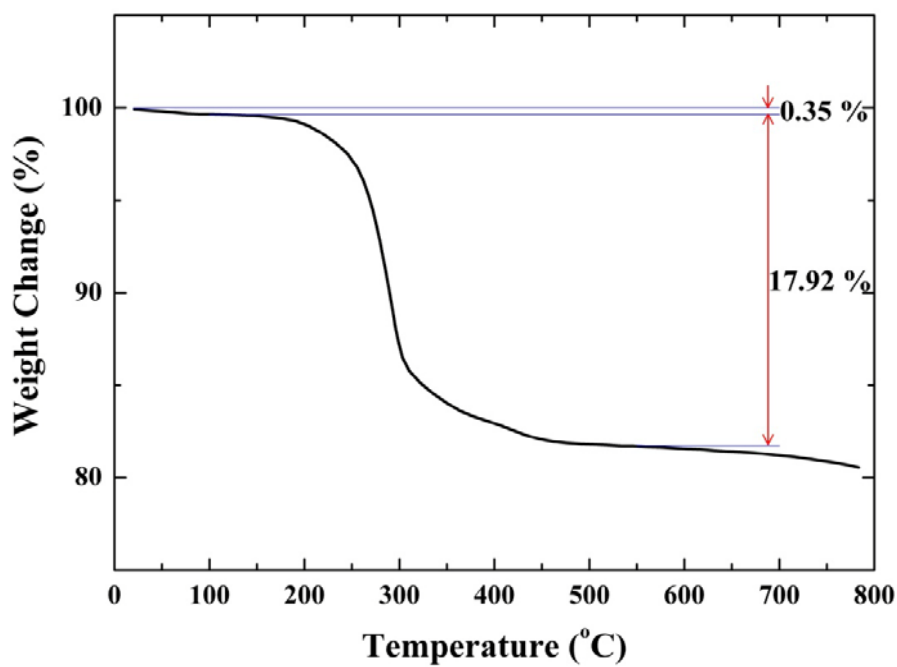

**Fig. S4: TGA curve of as-synthesized silver nanoparticles.** 0.35% weight loss is due to residual water. 17.92% decrease in weight is due to the loss of organic capping material (palmitic acid) and silver amount corresponds to 81.73% in weight.

### II.2. Inductively coupled plasma (ICP)-atomic emission spectrometry (AES)

The result of ICP-AES analysis is summarized as,

Ag nanoparticles (0.2 g) in 60 % Nitric acid (10 mL) = 20 g/L Ag nanoparticle solution

$Fe = 3.38 \times 10^{-7}$ g/L

$Si = 2.38 \times 10^{-6}$ g/L

As a result of TGA analysis, we found that a negligible quantity of Fe is contained in Ag nanoparticles ($1.69 \times 10^{-6}$ wt %).

## II.3. X-ray Photoemission Spectroscopy (XPS)

The XPS was performed to analyze elemental composition and to check impurity in the Ag nanoparticles. In the survey spectra, peaks originating from Ag, O, and C are identified. We could not find noticeable peaks from the magnetic elements, such as Fe, Co, and Ni. The existence of carbon and oxygen in the capping material (palmitic acid) explains the salient peaks of corresponding elements in the survey spectra (Fig. S5a). The Ag 3d peak in the survey spectra is composed of two complex peaks. The Ag $3d_{5/2}$ and $3d_{3/2}$ peaks are deconvoluted into two curves for Ag and AgO, respectively (Fig. S5b). With this spectrum, the existence of silver oxide is confirmed but the exact valence state of the silver ion could not be determined because rigorous quantitative analysis could not be performed due to a various silver oxide composition.

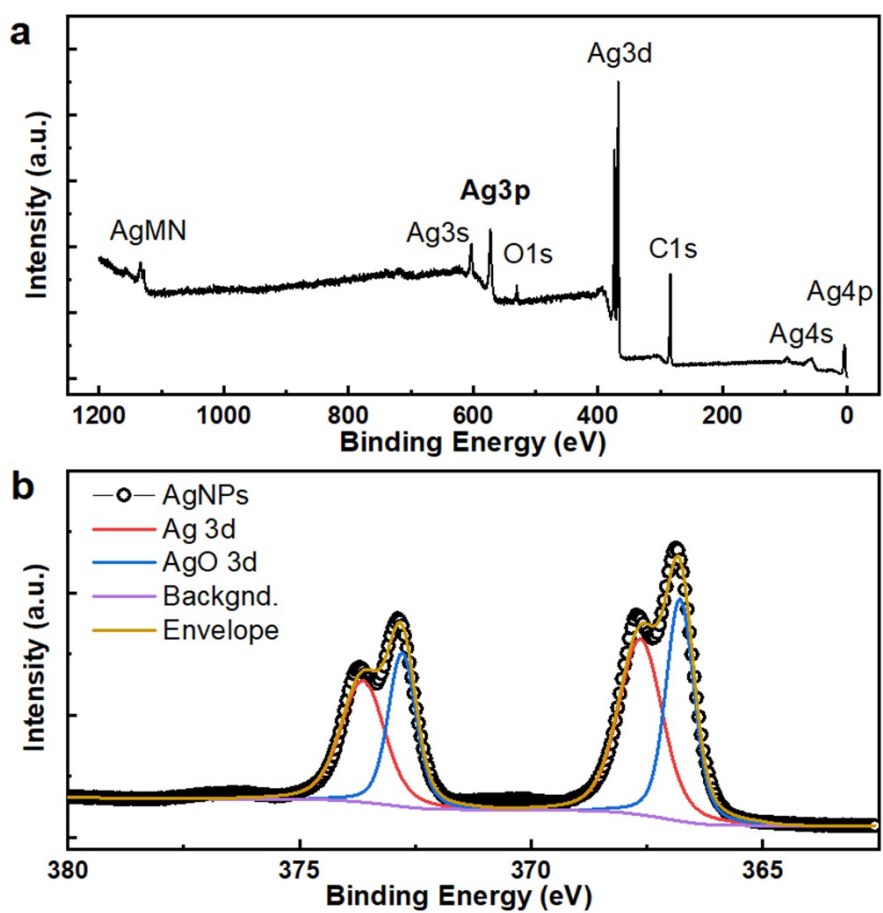

**Fig. S5: X-ray photoemission spectroscopy. a**, The survey and **b**, Ag 3d XPS spectra of Ag nanoparticles.

## III. Magnetization analysis

### III.1. Magnetization analysis

**Magnetization measurement**

The magnetic property of AgNP has been investigated with a sample in powder form. To perform SQUID measurements with powder sample, a container is required. Among the various types of containers, a gelatin capsule is found to be the most suitable for small signal samples with low susceptibility of about $10^{-7}$ emu/(g·Oe).[29] However, to obtain the magnetization of AgNP, the contribution of the capsule must be excluded. The compensation was done by performing an independent preparatory measurement of the empty gelatin capsule. As expected, the empty capsule exhibits diamagnetic behavior as is observed in the temperature dependent magnetization measurement (Fig. S6). Then, the AgNP specimen was put into the pre-measured capsule and the magnetization of the whole sample was measured. The magnetic moments of the AgNP sample and the capsule were measured together and the capsule contribution was subtracted to extract the "AgNP-only-sample" signal. The comparison of the total moment, measured capsule moment, and the extracted AgNP sample moment are shown in Fig. S6. The step observed around 25 K is an instrumental artifact. Such steps occur when the SQUID sensor signal polarity changes. It has nothing to do with the material properties.

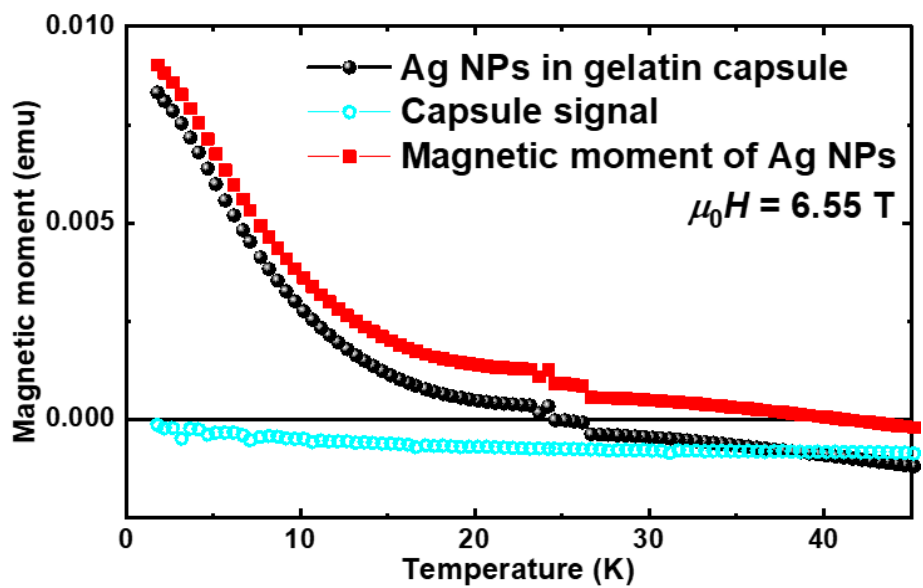

**Fig. S6: Raw data of magnetic moment vs. temperature curves at 6.55 T.** The AgNPs moment (red) was obtained by subtracting the empty capsule moment (cyan) from the total moment (black)

**Background signals**

Usually, a capping material is employed to prevent the agglomeration of the nanoparticles. However, the capping material also has its own magnetic signal. For the proper analysis, the magnetization of the capping material has to be checked independently. In the measurement of the capping material, a gelatin capsule was used in the same way as in the measurements of the AgNP sample. The empty capsule was measured first, and the capping material was put into the measured capsule and measured again. The diamagnetic susceptibility of the capping material was obtained by removing the capsule contribution. (Fig. S7) Deduced susceptibility of the capping material is $-7.35 \times 10^{-7}$ emu/(g·Oe). The value is similar to the theoretical value of the susceptibility of the capping material (palmitic acid) obtained by Pascal's constants method.[30] The diamagnetic susceptibility value for palmitic acid is estimated to be $-203.96 \times 10^{-6}$ emu/(mol·Oe) = $-7.95 \times 10^{-7}$ emu/ (gOe) by Pascal's constants method.

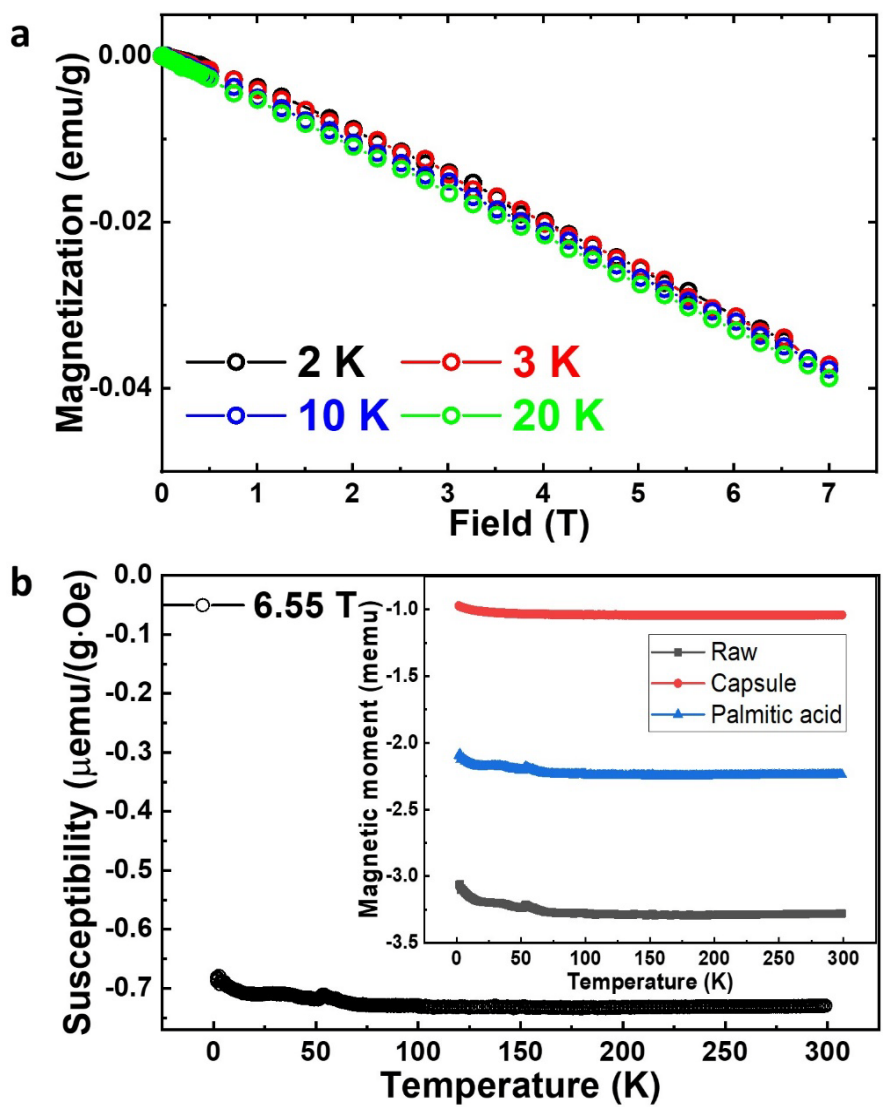

**Fig. S7: Magnetic behavior of the capping material. a**, The field-dependent magnetization at various temperature. **b**, The temperature-dependent magnetic susceptibility at 6.55 T. (inset: measured magnetic moment of whole sample (black), measured magnetic moment of "capsule-only" sample (red), magnetic moment of palmitic acid obtained by removing the empty capsule signal (blue).

**Another background**

By subtracting the contribution of the container, the magnetization of the pure AgNP is measured. The results of ICP-AES and XPS excluded the possibility of the contribution of the magnetic impurities such as Fe, Co, etc. in the measured magnetization of the AgNP.

As previously described, the magnetization of AgNP below 30 K can be explained in terms of electron number parity in nanometre-scale confinement. However, when the FC (Field Cooled) magnetization of AgNP was measured over an extended temperature range, a step-like behavior was observed between 40 K and 80 K. (Fig. S8) A similar step-like behavior in FC magnetization of silver nanoparticles produced by the decomposition of silver oxalate has been previously reported.[31] Interestingly, ZFC (Zero Field Cooled) magnetization data and FC magnetization data in the paper bifurcate at the FC magnetization step, suggesting that there is a "ferromagnetic-like" component in the AgNP sample. On the other hand, a "paramagnetic-like" rise is observed in the low-temperature region in both ZFC and FC magnetization data. This phenomenon implies that the low-temperature "paramagnetic-like" magnetization rise is independent of the "ferromagnetic-like" behavior.

It is noteworthy that there was a report on the ferromagnetic properties of silver oxide.[32] In the paper, it is proposed that the non-stoichiometric oxidation of silver exhibits ferromagnetic behavior. Since the presence of silver oxide in the AgNP specimens was confirmed by XPS measurements, we speculate that the step in the FC magnetization data and the ZFC-FC bifurcation are due to the non-stoichiometric oxidation of silver in the AgNP core.

The presence of such "ferromagnetic-like" components does not change the NMR results and their analysis. In general, ferromagnetic components cause large shifts and/or broadening of the spectrum. It is argued that the influence of the "ferromagnetic-like" component on the NMR results of this study is negligible because no excessive spectral broadening was observed and only the "un-shifted" NMR signal was measured and analyzed.

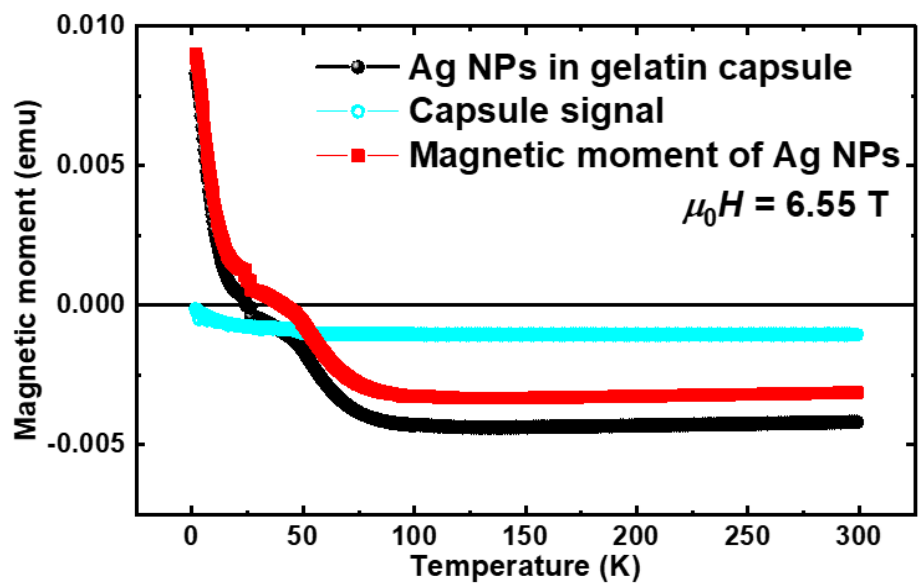

**Fig. S8:** Extended temperature range MT curves of AgNP, capsule, and AgNP in gelatin capsule measured at 6.55 T.

# IV. Supporting information for NMR measurement

## IV.1. Wide band solid state nuclear magnetic resonance experiment

It is worth mentioning that NMR is one of the most utilized techniques for investigating spin dynamics that involves magnetic and non-magnetic degrees of freedom, thanks to the use of a local probe (the nucleus) that detects spin dynamics not reachable by means of macroscopic tools, like *e.g.* AC and DC magnetometry, specific heat and so on. NMR has the ability to unravel the local magnetic dynamics in a wide range of wave-vectors and frequencies, the latest ones typically going from some MHz to several hundredths of MHz (extensible with special devices to the kHz and the GHz regions). On the other hand, in their review[33] Van der Klink and Brom showed how several authors, in metal clusters/particles, using the NMR resonance line position and the nuclear relaxation rates, determined the electronic states and the density of states at the Fermi level, in addition to the character (e.g. s or d) of the different bands.

The local field fluctuation at the proton sites in AgNP was investigated with $^1$H solid state NMR to study the spin fluctuations in the silver nanoparticle cores. (Silver nuclei were not probed because of very long spin-lattice relaxation time and very weak NMR sensitivity.) We also performed $^1$H NMR experiment on the "capping material only" sample for comparison.

## IV.2. Temperature and field dependent $^1$H NMR spectrum linewidth

In Fig. S9, we show the linewidth (FWHM, Full Width at Half Maximum) vs. temperature data measured at two different fields of 1.299 T and 3 T. In both sets of data, the palmitic acid linewidth data show a almost mild increase with decreasing temperature and an increase with increasing field. The temperature dependence of linewidth of the palmitic acid spectrum can be understood in terms of thermal motion within a rigid lattice model. The thermally activated motion of part of or whole palmitic acid changes the dipolar interaction tensor between the nuclei (protons). This, in turn, alters the linewidth of the Gaussian spectra. It is worthwhile to mention that similar thermal motion related linewidth change in the $CH_3$ group has been previously reported.[34] Also, we attribute the field dependence of the linewidth to the magnet inhomogeneity related field spread ($\Delta H_0$) which is field ($H_0$) dependent. (If one assumes constant field inhomogeneity, $\Delta H_0/H_0 =$ constant, then $\Delta H_0 \propto H_0$.)

Putting all these things together, we devised an empirical equation for the analysis. Experimental data of palmitic acid linewidth is fitted with the empirical equation,

$$\text{Linewidth (FWHM)} = \gamma \alpha H + [B - AT]. \tag{1}$$

In the equation, the first term is a field-dependent but temperature-independent term which describes field inhomogeneity induced line broadening. In the term, $\alpha$ is defined as temperature-independent field inhomogeneity constant, $\alpha \equiv \Delta H_0/H_0$, and $\gamma$ is the proton gyromagnetic ratio. The second term is included to take into account the temperature dependent linewidth behavior due to molecular motion. As a simplest approximation, we assumed the linear temperature dependence. The fitting has been done on two sets of palmitic acid data (1.299 T and 3 T) and two sets of high temperature AgNP data (T > 150 K, 1.299 T and 3 T). The high temperature AgNP data sets are included in the fitting because of the following reason; it is noted that the high temperature AgNP linewidth data is very similar to the high temperature palmitic acid linewidth data. The similarity is physically reasonable because the magnetization of the AgNP core is diminished to very small value at high temperature. Recalling that the electron magnetic moment in AgNP core contribute to the local field at proton sites in AgNP via dipolar interaction, we can say that the influence of the AgNP core on the linewidth is negligibly small at high temperature.

The values of the three fitting parameters deduced are $\alpha = 2.06 \times 10^{-4}$, $B = 0.0534$ MHz, $A = 1.055 \times 10^{-4}$ MHz/K. It is noted that the deduced value of field inhomogeneity constant $\alpha$ is quite reasonable for field sweeping magnets. The resultant fitting curves are shown in the Fig. S9 as solid straight lines. (Green for 3 T data and cyan for 1.299 T data)

After the fitting, one interesting feature of the temperature dependence of AgNP linewidth draws attention: the linewidth of the AgNP spectrum increases more steeply than that of the palmitic acid below 100 K as the temperature is decreased. The feature is more evident in the 3 T data. Remarkably, such a behavior of the linewidth is reminiscent of the magnetization expected for odd electron-number parity confinements. To be specific, the magnetization of AgNP is temperature-independent regardless of the electron-number parity for temperatures above $\delta/k_B$. (where $\delta$ is the average level spacing of the discrete energy level of AgNP) Below around $T = \delta/k_B$, the odd-parity magnetization rises rapidly with decreasing temperature.[9] Such a similarity is plausible from a physical perspective. As mentioned before, the dominant source of the local field at proton sites in AgNP is the dipolar field exerted by the electron magnetic moments in AgNP core. Then, the linewidth of the spectra is determined by the distribution width of the dipolar fields at the proton sites, which can be related to the source strength, i.e., magnetization, in conjunction with the randomness of the dipolar interaction tensor. Thus, on a qualitative level, the behavior of the AgNP spectrum linewidth is compatible with the electron-number parity scenario.

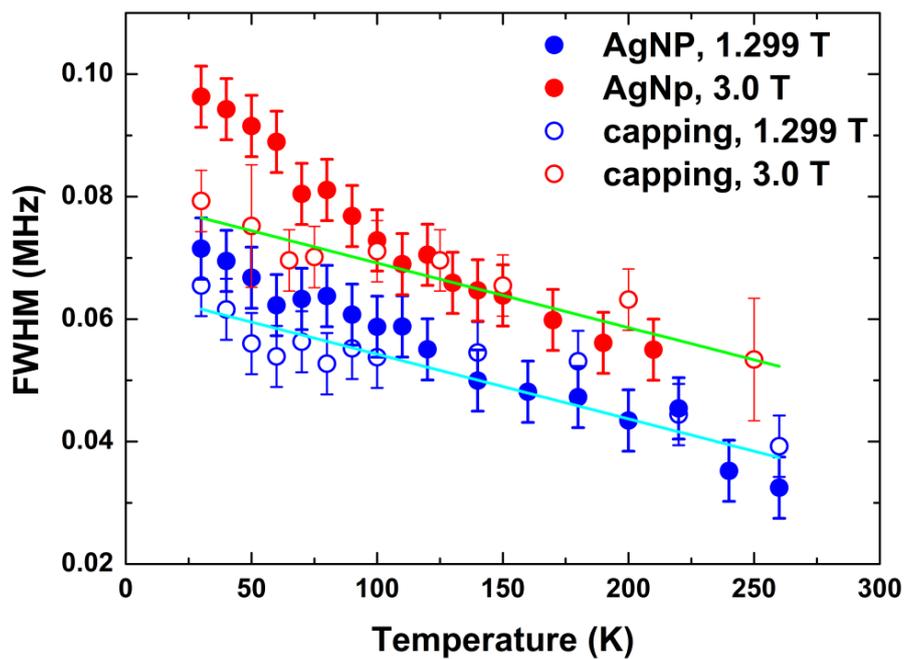

**Fig. S9:** The linewidth (FWHM) vs. temperature for AgNP and palmitic acid measured at 1.299 T and 3.0 T

## IV.3. $^1$H NMR spin-spin relaxation ($T_2$) measurement

Spin-spin relaxation time $T_2$'s of AgNP and capping material were measured as functions of temperature at 1.299 T (both AgNP and capping material) and 3.0 T (AgNP only). Below 150 K, measured $T_2$'s of AgNP and palmitic acid are almost the same and almost temperature independent. Thus we can conclude that the influence of Ag nanoparticle cores on $^1$H NMR spin-spin relaxation is quite negligible in the temperature range. Above 150 K, a slight temperature dependence is noticed and the trend is quite clear in the data of AgNP. (see Fig. S10.)

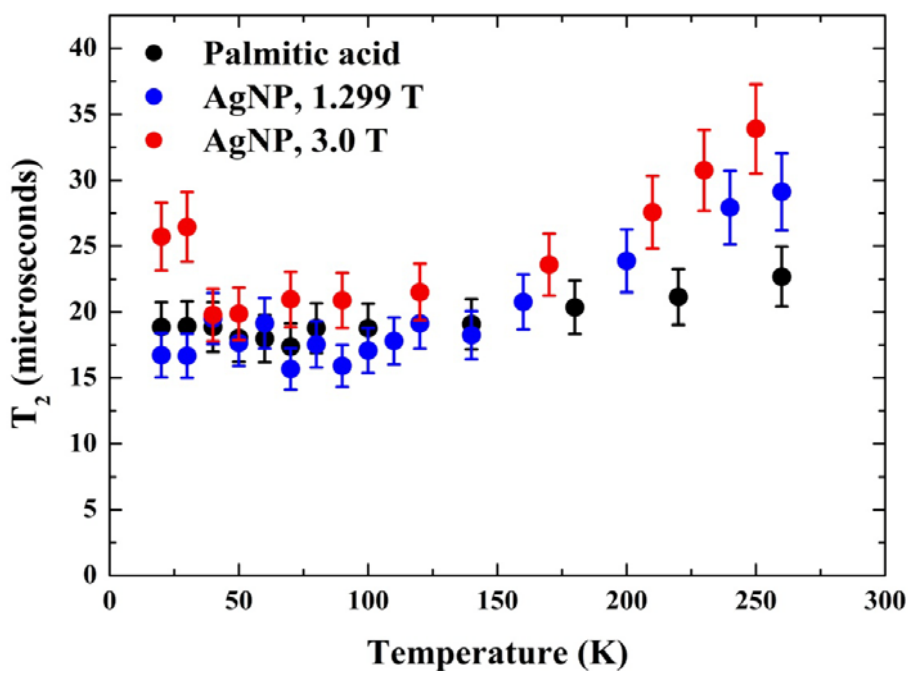

**Fig. S10: Temperature dependence of $T_2$ of AgNP and palmitic acid measured at 1.299 T and 3.0 T. Palmitic acid $T_2$ was measured at 1.299 T.**

## IV.4. $^1$H NMR nuclear spin-lattice relaxation ($T_1$) measurement

$^1$H NMR nuclear spin-lattice relaxation ($T_1$) behavior was investigated at two fields (1.299 T and 3.0 T) as functions of temperature. As a monitoring sequence, the Hahn echo sequence ($\pi/2 - \pi$) is mainly used and, depending on the RF signal conditions, the solid echo sequence ($\pi/2 - \pi/2$) is also utilized intermittently. We checked the difference in recovery behavior between the measurements with Hahn echo and the measurements with solid echo. Within the experimental error bound, the two data sets agree well with each other. (see Fig. S11.)

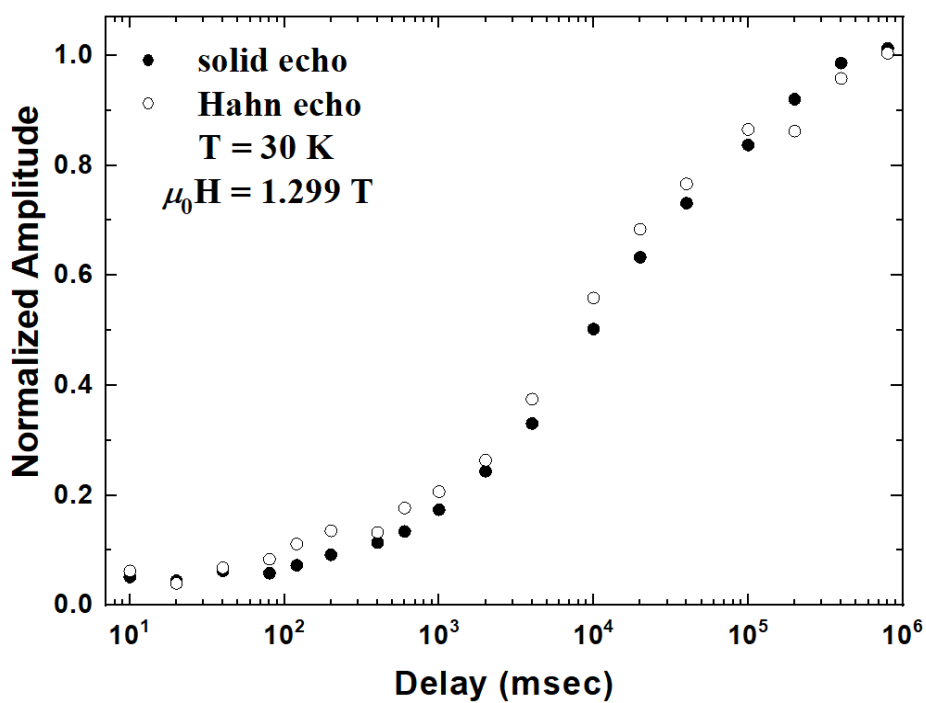

**Fig. S11: Comparison of the nuclear magnetization recovery curves measured with Hahn echo and solid echo as monitoring pulse sequences.**

# V. Supporting information for the theory on the nuclear spin-lattice relaxation

## V.1. Gaussian modulation approximation

In order to describe the nuclear spin-lattice relaxation in a magnetic system in presence of correlated spin dynamics, it is more convenient to express the nuclear spin-lattice relaxation rate $1/T_1$ in terms of the response functions by using the fluctuation-dissipation theorem[21, 35]:

$$\frac{1}{T_1} = \frac{(\hbar\gamma_e\gamma_n)^2}{4\pi g^2 \mu_B^2} k_B T \left[ \frac{1}{4} \sum_{\vec{k}} A^{\pm}(\vec{k}) \chi^{\pm}(\vec{k}) S^{\pm}_{\vec{k}}(\omega_e) + \sum_{\vec{k}} A^z(\vec{k}) \chi^z(\vec{k}) S^z_{\vec{k}}(\omega_n) \right], \quad (2)$$

where $\gamma_n$ and $\gamma_e$ are the gyromagnetic ratios of the nucleus and the electron, respectively. The coefficients $A^{\pm}(\vec{k})$ and $A^z(\vec{k})$ are the Fourier transforms of the spherical components of the product of two dipole-interaction tensors which describe the hyperfine coupling of a given proton to the electron spin moments.[36] In the coefficients, the symbols $\pm$ and z refer to the transverse and longitudinal components with respect to the quantization direction which is the direction of the external magnetic field. The collective $\vec{k}$-dependent spin correlation function is written as the product of the static response function (susceptibility) and the normalized relaxation function $f^{\pm,z}_{\vec{k}}(t)$. $S^{\pm,z}_{\vec{k}}(\omega)$ in Eq.(2) is the spectral density function, which is the Fourier transform of the normalized relaxation function. All other constants are defined as usual.[21, 22, 36]

In the classical limit where $k_B T \gg E$ ($E$ is the magnetic interaction energy between electron spins), one can neglect the $\vec{k}$-dependence of the generalized susceptibility $\chi^{\pm,z}(\vec{k})$ and the spectral density function, $S^{\pm,z}_{\vec{k}}(\omega)$.[21] Furthermore, with the assumption of isotropic susceptibility, one can safely set as $(1/2)\chi^{\pm} = \chi^z = \chi(\vec{k} = 0)$. It is noted that $\chi(\vec{k} = 0)$ is the static susceptibility which we can calculate with the scheme described in Ref. [9]. Incorporating all the notions, we get,

$$\frac{1}{T_1} = \frac{(\hbar\gamma_e\gamma_n)^2}{4\pi g^2 \mu_B^2} k_B T \chi(\vec{k}=0) \left[\frac{1}{2}\mathcal{S}^{\pm}(\omega_e)\sum_{\vec{k}} A^{\pm}(\vec{k}) + \mathcal{S}^z(\omega_n)\sum_{\vec{k}} A^z(\vec{k})\right]. \quad (3)$$

$\vec{k}$-space summation can be approximated by the product of *the number of terms in the summation* and *the $\vec{k}$-independent average value for the dipolar interaction tensors, such as* $\overline{A^{\pm}(\vec{k})} \Rightarrow A^{\perp}$, $\overline{A^z(\vec{k})} \Rightarrow A^{\parallel}$.[21] If there are $N$ electrons which interact with a given proton, the $\vec{k}$-space summation should have $N$ terms. Therefore, Eq. (3) is reduced to,

$$\frac{1}{T_1} = \frac{N(\hbar\gamma_e\gamma_n)^2}{4\pi g^2 \mu_B^2} k_B T \chi(\vec{k}=0) \left[\frac{1}{2}A^{\perp}\mathcal{S}^{\pm}(\omega_e) + A^{\parallel}\mathcal{S}^z(\omega_n)\right]. \quad (4)$$

Exact determination of the spectral density function or, equivalently, the normalized relaxation function is not possible at this stage. Instead, we adopt an approximation which is in accordance with the experimental observations. Gaussian type spectrum of AgNP $^1$H NMR and negligibly small exchange interaction between the protons in AgNP (compared to the dipolar interaction) enable us to assume that the local field fluctuation correlation function at proton sites has a Gaussian type time dependence - Gaussian type normalized relaxation function.[23-25] In contrast to the palmitic acid case, the dominant contribution to the local field fluctuation at the proton sites in AgNP is the dipolar field of the electronic moments in AgNP core and the contribution makes the measured nuclear spin-lattice relaxation rate dramatically different from that of the palmitic acid. Therefore we conjecture that the time dependence of the correlation function of the electronic spins in AgNP is also a Gaussian, such as $f_{\vec{k}}^{\pm,z}(t) \sim \exp(-\Omega^2 t^2/2)$. (We also assumed isotropic normalized relaxation function.)

By performing Fourier transform of the Gaussian type normalized relaxation function, we can obtain the mathematical expression for the nuclear spin-lattice relaxation rate as,

$$\frac{1}{T_1} = \frac{N(\hbar\gamma_e\gamma_n)^2}{4\pi g^2 \mu_B^2} k_B T \chi(\vec{k}=0) \frac{\sqrt{2\pi}}{\Omega} \left[\frac{1}{2} A^\perp e^{-(\omega_e^2/\Omega^2)/2} + A^\parallel e^{-(\omega_n^2/\Omega^2)/2}\right]. \tag{5}$$

We assume that the spin-spin interaction energy between the "free" electrons in AgNP core is of the order of the dipolar interaction energy. Therefore, we set $\Omega$ in the expression as $(\varepsilon/\hbar)$ where $\varepsilon$ is the dipolar interaction energy between the electrons.

Proton sites in the capping material (palmitic acid) are grossly inequivalent. The site difference causes the continuous distribution of the $T_1$'s, resulting in stretched exponential type nuclear magnetization relaxation curve. It is worthwhile to note that $T_1^*$ values obtained from fitting the relaxation curves with $M(t) = M(t=\infty)\left[1 - \exp(-\{t/T_1^*\}^\beta)\right]$ are proportional to the most probable values in the $T_1$ distributions, and the proportionality constant is linearly dependent on the stretching exponent, $\beta$.[18] With the multiplication of the simple correction factor $B$, which is of the order of 1, $1/T_1^*$ can be approximated as,

$$\frac{1}{T_1^*} \approx \frac{BN(\hbar\gamma_e\gamma_n)^2}{4\pi g^2 \mu_B^2} k_B T \chi(\vec{k}=0) \frac{\sqrt{2\pi}}{\Omega} \left[\frac{1}{2} A^\perp e^{-(\omega_e^2/\Omega^2)/2} + A^\parallel e^{-(\omega_n^2/\Omega^2)/2}\right]. \tag{6}$$

Therefore, it is found that $1/T_1^*$ is proportional to the product of temperature and the static susceptibility $\chi(\vec{k}=0)$ such as, $1/T_1^* \propto T\chi(\vec{k}=0)$ and $1/T_1^* = \alpha T\chi(\vec{k}=0)$. By comparing Eq. (6) with the equation used in fitting the experimental data, $1/(T_1^* T) = \alpha(\chi_{np} + \eta)$, the expression for the proportionality constant $\alpha$ can be identified as,

$$\alpha = \frac{BN(\hbar\gamma_e\gamma_n)^2}{4\pi g^2 \mu_B^2} k_B \frac{\sqrt{2\pi}}{\Omega} \left[\frac{1}{2} A^\perp e^{-(\omega_e^2/\Omega^2)/2} + A^\parallel e^{-(\omega_n^2/\Omega^2)/2}\right]. \tag{7}$$

The value of $\alpha$ obtained from fitting the experimental data is $3.74 \times 10^{25}$ [Oe·s$^{-1}$·K$^{-1}$·emu$^{-1}$]. Using this value of $\alpha$ and proper values or estimated values of other physical quantities in the Eq. (7), we could assess the validity of our model by estimating the distance between an electron and a proton. The distance can be estimated by calculating the spherical components of the product of two dipole interaction tensors, $A^\perp$ and $A^\parallel$, which are inversely proportional to the sixth power of the distance.

First, we estimate the value of $\Omega$. Assuming the distance between the conduction electrons is about ~ 4 Å, $\Omega$ is calculated to be of the order of ~$1.2 \times 10^9$ [rad/s]. Using the electronic Larmor frequency ($\omega_e$) and the nuclear Larmor frequency ($\omega_n$) of the experiments, we find that $\Omega^{-1}e^{-(\omega_n/\Omega)^2/2} \gg \Omega^{-1}e^{-(\omega_e/\Omega)^2/2}$ and $\Omega^{-1}e^{-(\omega_n/\Omega)^2/2} \sim 7.99 \times 10^{-10}$ [rad/s]$^{-1}$. With these estimations and with the assumption that $A^\perp$ and $A^\parallel$ are of the same order of magnitude, Eq. (7) can be approximated as,

$$\alpha \cong \frac{BN(\hbar\gamma_e\gamma_n)^2}{4\pi g^2\mu_B^2} k_B \frac{\sqrt{2\pi}}{\Omega} A^\parallel e^{-\frac{\omega_n^2/\Omega^2}{2}}, \tag{8}$$

$$\frac{\alpha\sqrt{128\pi}}{BN\gamma_n^2 k_B}\Omega e^{(\omega_n^2/\Omega^2)/2} = \frac{7.591 \times 10^{33}}{BN}\Omega e^{(\omega_n^2/\Omega^2)/2} \cong A^\parallel \sim \frac{4.5}{r^6}. \tag{9}$$

For the estimation of the proton-electron distance, the correction factor $B$ for stretched exponential is set to be 0.5 as a rough estimation. The number of "free" electron $N$ is calculated to be ~ 1500 for 3 nm cube. By setting $N \sim 1500$, the distance between a proton and an electron is estimated to be ~ 2.7 nm. The number is in the right range considering the geometrical structure of AgNP, which indicates the validity of our model.

### V.2. Spin diffusion in hydrodynamic limit

Spin diffusion is an important mechanism for nuclear spin relaxation by electrons in solids.[19] In general, the spins of delocalized electrons are magnetically compensated in nonmagnetic metals. In the presence of odd electron number parity confinements, however, there are uncompensated electron spins. The uncompensated spins are expected to exist in odd electron-number parity confinements. In this case, the magnetic interaction between electrons leads to the spin diffusion, justifying the use of hydrodynamic description of spins. We start from the following formula for the calculation of relaxation time $T_1$[37] and make some approximation.

$$\frac{1}{T_1} = \frac{2\gamma^2 k_B T}{g^2 \mu_B^2} A_{\mathbf{hf}}^2 \sum_{\mathbf{q}} \frac{\chi_\perp''(\mathbf{q}, \omega_0)}{\omega_0}, \tag{10}$$

where $\gamma$ is the nuclear gyromagnetic ratio, $\omega_0$ is the nuclear resonance frequency, $\chi_\perp''(\mathbf{q}, \omega_0)$ is the imaginary part of the magnetic susceptibility component with wave vector $\mathbf{q}$ and frequency $\omega_0$ perpendicular to the direction of the external magnetic field, $A_{\mathbf{hf}}$ is the hyperfine coupling constant, $\mu_B$ is Bohr magneton, and $g$ is electron g factor. A hydrodynamic description is valid if the magnetization varies slowly in space and time.[38] For confinements of the size $d$ with the spin diffusion constant $D$, hydrodynamics can be used when $q$ is smaller than $\frac{2\pi}{d}$ and $\omega$ is smaller than $\frac{12\pi D}{d^2}$. Therefore, we have $\frac{2\pi}{q} > d$ for spatial average, and $\sqrt{6D \frac{2\pi}{\omega}} > d$ for diffusion equation to be valid for the spins in the confinement of the size $d$. In this case, we may write the following formula.[38-40]

$$\frac{\chi''(\mathbf{q}, \omega_0)}{\omega_0} = \chi(\mathbf{q}) \frac{Dq^2}{\omega^2 + (Dq^2)^2} \tag{11}$$

$$\chi_0 = \frac{\partial M}{\partial h} = \lim_{q \to 0} \chi(\mathbf{q}) \tag{12}$$

For the given value of $\omega$, $\frac{Dq^2}{\omega^2 + (Dq^2)^2}$ has maximum at $\omega = Dq^2$. Therefore, we may use the following approximation for the calculation of $T_1$ as long as $\chi\left(\sqrt{\frac{\omega_0}{D}}\right) \approx \chi_0$.

$$\frac{\chi_\perp''(\mathbf{q}, \omega_0)}{\omega_0} \approx \chi\left(\sqrt{\frac{\omega_0}{D}}\right) \frac{Dq^2}{\omega_0^2 + (Dq^2)^2} \approx \chi_0 \frac{Dq^2}{\omega_0^2 + (Dq^2)^2}. \tag{13}$$

$$\frac{1}{T_1} \approx \frac{2\gamma^2 k_B T}{g^2 \mu_B^2} \chi_0 A_{\mathbf{hf}}^2 \sum_{\mathbf{q}} \frac{Dq^2}{\omega_0^2 + (Dq^2)^2}. \tag{14}$$

The value of $D$ can be estimated from the Heisenberg paramagnet approximation. Bennett and Martin calculated the high-temperature limit of $D$ from the nonlinear integral equation for spin diffusion in the Heisenberg paramagnet and the spherical continuum model.[40] In the spherical continuum model, the high temperature limit value of $D$ is given as,

$$D \to \frac{j\lambda^2}{\hbar} \left[ \frac{S(S+1)}{3} \frac{v \tan^{-1} \kappa}{48\pi^2 \lambda^2} \right]^{1/2}, \tag{15}$$

where

$$S(S+1) = \text{the magnitude of the spin,}$$

$$v = \text{the average volume per spin,}$$

$$\lambda = \text{the effective interaction range,}$$

$$\kappa = \lambda q_0,$$

$$\text{and } j = 4\pi J \lambda^2 / v.$$

Here, $J$ is the parameter for the magnetic interaction $J(\vec{r}) = J\lambda\left(e^{-(|\vec{r}|/\lambda)}/|\vec{r}|\right)$, with $\vec{r}$ the difference in site location, and $q_0 = (6\pi^2/v)^{1/3}$. Similar results were obtained with slightly different numerical factors.[40-42]

The spherical continuum model was proposed to simplify the lattice structure and exchange interaction.[40] Similarly, in the case of nanoparticles, the spherical continuum model is applicable to simplify the crystallographic structure in the domain of AgNP and magnetic interaction. $J$ is estimated to be smaller than thermal energy because of the observed paramagnetism. To take it into account that the volume of AgNP is finite, we use $q_0 = (6\pi^2/v + q_c^3)^{1/3}$, where $q_c = (6\pi^2/v_{AgNP})^{1/3}$ and $v_{AgNP}$ is the average volume per silver nanoparticle core.

We note that the high-temperature limit of $D = \frac{j\lambda^2}{\hbar}\left[\frac{S(S+1)}{3}\frac{v\tan^{-1}\kappa}{48\pi^2\lambda^2}\right]^{1/2}$, is obtained with the assumption that $\kappa \tan^{-1} \kappa \gg 1$. In case that $\kappa \tan^{-1} \kappa \gg 1$ does not hold, we may use $D = \frac{j\lambda^2}{\hbar}\left[\frac{S(S+1)}{3}\frac{v f(\kappa)}{48\pi^2\lambda^2}\right]^{1/2}$ with $f(\kappa) = 16\int_0^\kappa \frac{x^2}{(1+x^2)^4}dx$ as long as $J$ is sufficiently small.[40]

We may use the dipolar interaction in the spin diffusion formulation by using $J$ estimated from dipolar interaction. In the case of silver, we estimate $J = \frac{\mu_0}{4\pi}\frac{\mu_B^2}{a_{Ag}^3} \approx 1.26 \times 10^{-25}$ J, where $a_{Ag}$ is the lattice constant of the conventional unit cell of silver and $\mu_0$ is vacuum magnetic permeability. Using $d = 2.4$ nm for the diameter of Ag domain, $d_{AgNP} = 7.0$ nm for the diameter of silver nanoparticle core, $\lambda = a_{Ag}$, $v = d^3$, and $v_{AgNP} = d_{AgNP}^3$, we estimate $D \approx 1.65 \times 10^{-7}$ m$^2$s$^{-1}$. Therefore, we obtain $\frac{2\pi}{d} \approx 2.62 \times 10^7$ cm$^{-1}$ and $\frac{12\pi D}{d^2} \approx 1.08 \times 10^{12}$ s$^{-1}$. With $\omega_0 = 2\pi \times 127$ MHz $= 7.98 \times 10^8$ rad s$^{-1}$, we obtain $\sqrt{\frac{\omega_0}{D}} \approx 6.94 \times 10^5$ cm$^{-1}$. It is clear that $\sqrt{6D\frac{2\pi}{\omega_0}} > d$ and $\frac{2\pi}{\sqrt{\frac{\omega_0}{D}}} > d$, which justifies the hydrodynamic description. The expression for $\frac{1}{T_1}$ can be conveniently written as

$$\frac{1}{T_1 T} \approx \alpha \chi_0. \tag{16}$$

Here, $\alpha = \frac{2\gamma^2 k_B}{g^2 \mu_B^2} A_{hf}^2 \sum_q \frac{Dq^2}{\omega_0^2 + (Dq^2)^2}$. We estimate $\alpha = 5.77 \times 10^{29}$ m$^{-3}$ K$^{-1}$ s$^{-1}$ $= 7.26 \times 10^{24}$ Oe s$^{-1}$ K$^{-1}$ emu$^{-1}$, with $\gamma = 2.6752218744 \times 10^8$ rad T$^{-1}$s$^{-1}$ and $A_{hf}^2 = 9.09$ J m$^{-3}$. $A_{hf}^2$ is set to be $J$ divided by the volume of silver nanoparticle core, and $\gamma = 2.6752218744 \times 10^8$ T$^{-1}$ s$^{-1}$ is the proton gyromagnetic ratio $\gamma_p = 2\mu_p/\hbar$.